\documentclass[a4paper,aps,twocolumn,nofootinbib]{revtex4}
\RequirePackage[colorlinks,hyperindex]{hyperref}
\RequirePackage[english]{babel}
\RequirePackage[latin1]{inputenc}
\RequirePackage[T1]{fontenc}
\RequirePackage{mathrsfs}
\RequirePackage{amsmath}
\RequirePackage{amssymb}
\RequirePackage{amsbsy}
\RequirePackage{color}
\RequirePackage{bm}
\hypersetup{colorlinks=true,breaklinks=true,urlcolor=blue,linkcolor=red}
\begin{document}
\title{\bf{Polar solutions with tensorial connection of the spinor equation}}
\author{Luca Fabbri}
\affiliation{DIME, Sez. Metodi e Modelli Matematici, Universit\`{a} di Genova, 
Via all'Opera Pia 15, 16145 Genova, ITALY}
\date{\today}
\begin{abstract}
Dirac field equations are studied for spinor fields without any external interaction and when they are considered on a background having a tensorial connection with a specific non-vanishing structure some solution can be found in polar form displaying a square-integrable localized behaviour.
\end{abstract}
\maketitle
\section{Introduction}
Physics as seen from a very general perspective consists in writing a system of field equations and then finding the corresponding solutions. Once found, these solutions are applied to particular cases so to make specific predictions that are later compared to experiments and observations.

Because finding solutions is a very difficult endeavour, it has become customary to make predictions by exploiting properties that do not require having the explicit form of a special solution. Yet, the quest for solutions, though not central in contemporary fashion, is still fundamental in general because explicit solutions do have a complete information content. Just the same, due to the difficulty of the task, explicit solutions are rare, and they generally possess properties that may be undesirable. For example, in quantum field theory explicit solutions are in form of plane waves, which are not square-integrable, hence their conserved quantities are infinite and this is to be regarded as giving rise to meaningless interpretations of results.

Dirac fields \cite{G, L, Cavalcanti:2014wia, Fabbri:2016msm, h1, Fabbri:2016laz, Fabbri:2016fxt, Obukhov:2017avp} are one of the most important fields in nature, but so far as we are aware, there is no explicit solution that has been found and which has the desirable property of square-integrability. To be precise, there are explicit solutions that are square-integrable, such as the electronic orbitals in a Coulomb potential. Nevertheless, these solutions are based on external fields that are kept fixed, so they do not describe a complete solution, like a spinor and its surrounding electrodynamic field as given in the form of simultaneous solutions of the fully-coupled system of the Dirac and the Maxwell field equations.

And in any case, we known no solution for spinors having no external interactions, or even self-interaction, and that is in the free situation. However, by taking considerable advantage of the methods that we have drawn in references \cite{Fabbri:2017pwp, Fabbri:2017xyk, Fabbri:2018crr}, some solutions, explicit, square-integrable and completely self-sustained, could be obtained.

This is what we will be showing in this work.
\section{Geometry of spinors}
We shall consider a geometry having both torsion and curvature as background for the spinor fields \cite{G}, and as a start we recall that $\boldsymbol{\gamma}^{a}$ are the Clifford matrices, from which $\left[\boldsymbol{\gamma}_{a}\!,\!\boldsymbol{\gamma}_{b}\right]
\!=\!4\boldsymbol{\sigma}_{ab}$ and $2i\boldsymbol{\sigma}_{ab}
\!=\!\varepsilon_{abcd}\boldsymbol{\pi}\boldsymbol{\sigma}^{cd}$ are the definitions of the $\boldsymbol{\sigma}_{ab}$ and the $\boldsymbol{\pi}$ matrix (this matrix is what is usually indicated as gamma with an index five, but here we prefer to employ a notation with no index because the appearance of any meaningless index might be source of a number of misunderstandings for some readers).

As it was discussed \cite{L,Cavalcanti:2014wia,Fabbri:2016msm}, we have that in general, any spinor field can be written according to the form
\begin{eqnarray}
&\!\psi\!=\!\phi e^{-\frac{i}{2}\beta\boldsymbol{\pi}}
\boldsymbol{S}\left(\!\begin{tabular}{c}
$1$\\
$0$\\
$1$\\
$0$
\end{tabular}\!\right)
\label{spinor}
\end{eqnarray}
with $\boldsymbol{S}$ any complex Lorentz transformation: this form is what is normally called polar form of the spinorial field.

With a spinor and its conjugate spinor, we define the bilinear axial-vector and vector given according to
\begin{eqnarray}
&S^{a}\!=\!\overline{\psi}\boldsymbol{\gamma}^{a}\boldsymbol{\pi}\psi\!=\!2\phi^{2}s^{a}\\
&U^{a}\!=\!\overline{\psi}\boldsymbol{\gamma}^{a}\psi\!=\!2\phi^{2}u^{a}
\end{eqnarray}
and the bilinear pseudo-scalar and scalar as
\begin{eqnarray}
&\Theta\!=\!i\overline{\psi}\boldsymbol{\pi}\psi\!=\!2\phi^{2}\sin{\beta}\label{b2}\\
&\Phi\!=\!\overline{\psi}\psi\!=\!2\phi^{2}\cos{\beta}\label{b1}
\end{eqnarray}
with $u^{a}$ the velocity vector and $s^{a}$ the spin axial-vector and where $\phi$ is a scalar and $\beta$ is a pseudo-scalar known as module and Yvon-Takabayashi angle (we stress that the name Takabayashi can sometimes be spelled Takabayasi).

One can easily prove that the directions are such that
\begin{eqnarray}
&u_{a}u^{a}\!=\!-s_{a}s^{a}\!=\!1\\
&u_{a}s^{a}\!=\!0
\end{eqnarray}
with velocity having only the time component and spin having only the third component when in the polar form matrix $\boldsymbol{S}$ is the identity: module and Yvon-Takabayashi angle are thus the only two real degrees of freedom.

Computing the derivative of the spinor (\ref{spinor}), and since
\begin{eqnarray}
&\boldsymbol{S}\partial_{\mu}\boldsymbol{S}^{-1}\!=\!i\partial_{\mu}\theta\mathbb{I}
\!+\!\frac{1}{2}\partial_{\mu}\theta_{ij}\boldsymbol{\sigma}^{ij}\label{Lorentz}
\end{eqnarray}
where $\theta$ is a generic complex phase and $\theta_{ij}\!=\!-\theta_{ji}$ are the six parameters of the Lorentz group, we can define
\begin{eqnarray}
&\partial_{\mu}\theta_{ij}\!-\!\Omega_{ij\mu}\!\equiv\!R_{ij\mu}\label{R}\\
&\partial_{\mu}\theta\!-\!qA_{\mu}\!\equiv\!P_{\mu}\label{P}
\end{eqnarray}
which contain all information of the connection but are proven to be real tensors, therefore called tensorial connections: with them, the spinorial derivative becomes
\begin{eqnarray}
&\boldsymbol{\nabla}_{\mu}\psi\!=\!(\nabla_{\mu}\ln{\phi}\mathbb{I}
\!-\!\frac{i}{2}\nabla_{\mu}\beta\boldsymbol{\pi}
\!-\!iP_{\mu}\mathbb{I}\!-\!\frac{1}{2}R_{ij\mu}\boldsymbol{\sigma}^{ij})\psi
\label{decspinder}
\end{eqnarray}
from which we can get
\begin{eqnarray}
&\nabla_{\mu}s_{i}\!=\!R_{ji\mu}s^{j}\label{ds}\\
&\nabla_{\mu}u_{i}\!=\!R_{ji\mu}u^{j}\label{du}
\end{eqnarray}
identically; additionally
\begin{eqnarray}
\nonumber
&\!\!\!\!\!\!\!\!R^{i}_{\phantom{i}j\mu\nu}\!=\!\partial_{\mu}\Omega^{i}_{\phantom{i}j\nu}
\!-\!\partial_{\nu}\Omega^{i}_{\phantom{i}j\mu}
\!+\!\Omega^{i}_{\phantom{i}k\mu}\Omega^{k}_{\phantom{k}j\nu}
\!-\!\Omega^{i}_{\phantom{i}k\nu}\Omega^{k}_{\phantom{k}j\mu}=\\
&\!\!\!\!=-(\nabla_{\mu}R^{i}_{\phantom{i}j\nu}
\!-\!\!\nabla_{\nu}R^{i}_{\phantom{i}j\mu}\!+\!R^{i}_{\phantom{i}k\mu}R^{k}_{\phantom{k}j\nu}
\!-\!R^{i}_{\phantom{i}k\nu}R^{k}_{\phantom{k}j\mu})\label{Riemann}\\
\!\!&qF_{\mu\nu}\!=\!q(\partial_{\mu}A_{\nu}\!-\!\partial_{\nu}A_{\mu})
\!=\!-(\nabla_{\mu}P_{\nu}\!-\!\nabla_{\nu}P_{\mu})\label{Maxwell}
\end{eqnarray}
so that the parameters defined in (\ref{Lorentz}) have no curvature.

The Dirac equation in polar form is equivalent to
\begin{eqnarray}
\nonumber
&\frac{1}{2}\varepsilon_{\mu\alpha\nu\iota}R^{\alpha\nu\iota}
\!-\!2P^{\iota}u_{[\iota}s_{\mu]}+\\
&+2(\nabla\beta/2\!-\!XW)_{\mu}\!+\!2s_{\mu}m\cos{\beta}\!=\!0\label{dep1}\\
\nonumber
&R_{\mu a}^{\phantom{\mu a}a}
\!-\!2P^{\rho}u^{\nu}s^{\alpha}\varepsilon_{\mu\rho\nu\alpha}+\\
&+2s_{\mu}m\sin{\beta}\!+\!\nabla_{\mu}\ln{\phi^{2}}\!=\!0\label{dep2}
\end{eqnarray}
specifying all the first-order derivatives of the module and the YT angle \cite{h1,Fabbri:2016laz} with geometric field equations
\begin{eqnarray}
&\nabla^{2}P^{\mu}
\!-\!\nabla_{\sigma}\nabla^{\mu}P^{\sigma}\!=\!-2q^{2}\phi^{2}u^{\mu}\label{me}
\end{eqnarray}
alongside to
\begin{eqnarray}
&\!\!\!\!\nabla^{2}(XW)^{\mu}\!-\!\nabla_{\alpha}\nabla^{\mu}(XW)^{\alpha}
\!+\!M^{2}XW^{\mu}\!=\!2X^{2}\phi^{2}s^{\mu}\label{se}
\end{eqnarray}
as well as
\begin{eqnarray}
\nonumber
&\nabla_{k}R^{ka}_{\phantom{ka}a}g^{\rho\sigma}\!-\!\nabla_{i}R^{i\sigma\rho}
\!-\!\nabla^{\rho}R^{\sigma i}_{\phantom{\sigma i}i}\!+\!R_{ki}^{\phantom{ki}i}R^{k\sigma\rho}+\\
\nonumber
&+R_{ik}^{\phantom{ik}\rho}R^{k\sigma i}
\!-\!\frac{1}{2}R_{ki}^{\phantom{ki}i}R^{ka}_{\phantom{ka}a}g^{\rho\sigma}-\\
\nonumber
&-\frac{1}{2}R^{ika}R_{kai}g^{\rho\sigma}\!=\!\frac{1}{2}[M^{2}(W^{\rho}W^{\sigma}
\!\!-\!\!\frac{1}{2}W^{\alpha}W_{\alpha}g^{\rho\sigma})+\\
\nonumber
&+\frac{1}{4}(\partial W)^{2}g^{\rho\sigma}
\!-\!(\partial W)^{\sigma\alpha}(\partial W)^{\rho}_{\phantom{\rho}\alpha}+\\
\nonumber
&+\frac{1}{4}F^{2}g^{\rho\sigma}\!-\!F^{\rho\alpha}\!F^{\sigma}_{\phantom{\sigma}\alpha}-\\
\nonumber
&-\phi^{2}[(XW\!-\!\nabla\frac{\beta}{2})^{\sigma}s^{\rho}
\!+\!(XW\!-\!\nabla\frac{\beta}{2})^{\rho}s^{\sigma}-\\
\nonumber
&-P^{\sigma}u^{\rho}\!-\!P^{\rho}u^{\sigma}+\\
&+\frac{1}{4}R_{ij}^{\phantom{ij}\sigma}\varepsilon^{\rho ijk}s_{k}
\!+\!\frac{1}{4}R_{ij}^{\phantom{ij}\rho}\varepsilon^{\sigma ijk}s_{k}]]\label{ee}
\end{eqnarray}
specifying the second-order derivatives of gauge potentials, torsion and tetrad fields: as it is clear from expression (\ref{Riemann}), these are the field equations that determine the dynamics of electrodynamics, torsion and gravitation and their coupling to the current, spin and energy densities.

A general discussion on these field equations together with an attempt to find solitonic solutions can be found in reference \cite{Fabbri:2016fxt} while for some general treatment of various solutions we refer the reader to the recent reference \cite{Obukhov:2017avp}.

Extensive comments on the structure of regular as well as singular spinors may be found in references \cite{Fabbri:2017pwp, Fabbri:2017xyk}.

Explicit examples are provided in reference \cite{Fabbri:2018crr}.
\section{Restrictions and the case of the hydrogen atom}
In general, the above is a system of fully-coupled field equations that have to be simultaneously solved to obtain the complete solution; in practice, this is almost impossible: so we will begin by making some assumption.

A first assumption that comes to mind is to disregard the effects of torsion-gravity: torsion is so far unseen and gravity relatively weak, so it would make sense to search solutions without considering these effects; setting torsion to zero is given by $W_{a}\!=\!0$ but vanishing gravity is more difficult since the condition of flat space-time that is implemented by a null Riemann tensor $R^{i}_{\phantom{i}j\mu\nu}\!=\!0$ provides no further information. In \cite{Fabbri:2017pwp} and \cite{Fabbri:2017xyk} we have discussed how a reasonable assumption might be $R_{ijk}\!=\!0$ but in a further work \cite{Fabbri:2018crr} we also showed that such an assumption is untenable since the tensors $R_{ijk}$ can not equal zero in very general cases indeed: so there is no other way apart from finding a specific realization of the condition of vanishing Riemann tensor that does not involve any a priori assumption on the structure of the $R_{ijk}$ tensor. We have already mentioned that $R_{ijk}$ are a tensor containing the information about the reference system, which was called tensorial connection, and as such it encodes the information about a force that cannot be vanished despite being inertial, interpretable as a covariant type of inertial force.

However, one choice we can make is to choose spherical coordinates, in which the flat space-time may be described by the metric
\begin{eqnarray}
&g_{tt}\!=\!1\\
&g_{rr}\!=\!-1\\
&g_{\theta\theta}\!=\!-r^{2}\\
&g_{\varphi\varphi}\!=\!-r^{2}|\!\sin{\theta}|^{2}
\end{eqnarray}
with connection
\begin{eqnarray}
&\Lambda^{\theta}_{\theta r}\!=\!\frac{1}{r}\\
&\Lambda^{r}_{\theta\theta}\!=\!-r\\
&\Lambda^{\varphi}_{\varphi r}\!=\!\frac{1}{r}\\
&\Lambda^{r}_{\varphi\varphi}\!=\!-r|\!\sin{\theta}|^{2}\\
&\Lambda^{\varphi}_{\varphi\theta}\!=\!\cot{\theta}\\
&\Lambda^{\theta}_{\varphi\varphi}\!=\!-\cos{\theta}\sin{\theta}
\end{eqnarray}
and which give Riemann tensor equal to zero, as known.

Just the same, some restriction could be done, and as a start, we require restrictions on the velocity: specifically, we ask that $u_{k}$ have temporal and azimuthal components solely, so that by virtue of its normalization, they can be chosen according to 
\begin{eqnarray}
&u_{t}\!=\!\cosh{\alpha}\\
&u_{\varphi}\!=\!r\sin{\theta}\sinh{\alpha}
\end{eqnarray}
with $\alpha\!=\!\alpha(r,\theta)$ a generic function; orthogonality between velocity and spin allows us to consider $s_{k}$ having the other two components solely, and again by virtue of its property of normalization, we pick
\begin{eqnarray}
&s_{r}\!=\!\cos{\gamma}\\
&s_{\theta}\!=\!r\sin{\gamma}
\end{eqnarray}
with $\gamma\!=\!\gamma(r,\theta)$ a generic function. Because of (\ref{ds},\ref{du}) it now becomes possible to extract some information about the tensors $R_{ijk}$ in general: in fact, we may deduce that
\begin{eqnarray}
&R_{t\varphi\varphi}\!=\!R_{\varphi tt}\!=\!R_{r\theta\varphi}\!=\!R_{r\theta t}\!=\!0
\end{eqnarray}
as well as
\begin{eqnarray}
&r\sin{\theta}\partial_{\theta}\alpha\!=\!R_{t\varphi\theta}\\
&r\sin{\theta}\partial_{r}\alpha\!=\!R_{t\varphi r}\\
&r(1\!+\!\partial_{\theta}\gamma)\!=\!R_{\theta r\theta}\\
&r\partial_{r}\gamma\!=\!R_{\theta rr}
\end{eqnarray}
linking the derivatives of the two above functions to four of the components of the $R_{ijk}$ tensor and
\begin{eqnarray}
&\!rR_{rt\varphi}\!=\!R_{t\theta\varphi}\tan{\gamma}\\
&\!\!r\sin{\theta}R_{tr\varphi}\!=\!(R_{\varphi r\varphi}\!-\!r|\!\sin{\theta}|^{2})\tanh{\alpha}\\
&\!\!r\sin{\theta}R_{t\theta\varphi}\!=\!(R_{\varphi\theta\varphi}
\!-\!r^{2}\cos{\theta}\sin{\theta})\tanh{\alpha}\\
&\!\!\!\!-r(R_{\varphi r\varphi}\!-\!r|\!\sin{\theta}|^{2})
\!=\!(R_{\varphi\theta\varphi}\!-\!r^{2}\sin{\theta}\cos{\theta})\tan{\gamma}
\end{eqnarray}
as well as
\begin{eqnarray}
&rR_{rtt}\!=\!R_{t\theta t}\tan{\gamma}\\
&r\sin{\theta}R_{trt}\!=\!R_{\varphi rt}\tanh{\alpha}\\
&r\sin{\theta}R_{t\theta t}\!=\!R_{\varphi\theta t}\tanh{\alpha}\\
&rR_{r\varphi t}\!=\!R_{\varphi\theta t}\tan{\gamma}
\end{eqnarray}
and
\begin{eqnarray}
&rR_{rtr}\!=\!R_{t\theta r}\tan{\gamma}\\
&r\sin{\theta}R_{trr}\!=\!R_{\varphi rr}\tanh{\alpha}\\
&rR_{r\varphi r}\!=\!R_{\varphi\theta r}\tan{\gamma}\\
&r\sin{\theta}R_{t\theta r}\!=\!R_{\varphi\theta r}\tanh{\alpha}
\end{eqnarray}
with
\begin{eqnarray}
&rR_{rt\theta}\!=\!R_{t\theta\theta}\tan{\gamma}\\
&rR_{r\varphi\theta}\!=\!R_{\varphi\theta\theta}\tan{\gamma}\\
&r\sin{\theta}R_{t\theta\theta}\!=\!R_{\varphi\theta\theta}\tanh{\alpha}\\
&r\sin{\theta}R_{tr\theta}\!=\!R_{\varphi r\theta}\tanh{\alpha}
\end{eqnarray}
grouped in blocks. Notice that within each block the four components of the $R_{ijk}$ tensor are mutually related, but different blocks are independent on one another, so that it is possible to set an entire block to zero while leaving different from zero all components in any other block.

This allows us to have a certain care in setting some of the components equal to zero, and as an educated guess we can make the choice
\begin{eqnarray}
&R_{trr}\!=\!R_{t\theta r}\!=\!R_{\varphi rr}\!=\!R_{\varphi\theta r}\!=\!0\\
&R_{rt\theta}\!=\!R_{t\theta\theta}\!=\!R_{r\varphi\theta}\!=\!R_{\varphi\theta\theta}\!=\!0
\end{eqnarray}
as well as
\begin{eqnarray}
&R_{t\theta\varphi}\!=\!0\\
&R_{tr\varphi}\!=\!0
\end{eqnarray}
with
\begin{eqnarray}
&R_{r\varphi\varphi}\!=\!-r|\!\sin{\theta}|^{2}\\
&R_{\theta\varphi\varphi}\!=\!-r^{2}\cos{\theta}\sin{\theta}
\end{eqnarray}
and
\begin{eqnarray}
&R_{rtt}\!=\!-2\varepsilon\sinh{\alpha}\sin{\gamma}\\
&R_{\varphi rt}\!=\!2\varepsilon r\sin{\theta}\cosh{\alpha}\sin{\gamma}
\end{eqnarray}
with
\begin{eqnarray}
&R_{\theta tt}\!=\!2\varepsilon r\sinh{\alpha}\cos{\gamma}\\
&R_{\varphi\theta t}\!=\!-2\varepsilon r^{2}\sin{\theta}\cosh{\alpha}\cos{\gamma}
\end{eqnarray}
with $\varepsilon$ being a constant: notice that $\varepsilon$ can be interpreted as an integration constant, since these last relationships come from having imposed the vanishing of the Riemann curvature. This set of restrictions may look constraining, but we still retain the freedom to choose $\alpha$ and $\gamma$ and we still have the total freedom of choice for the $\varepsilon$ constant.

To propose an illuminating example, consider the case of the hydrogen atom in $1S$ orbital: in \cite{Fabbri:2018crr} we have seen that such an orbital is described on the background
\begin{eqnarray}
&R_{r\varphi\varphi}\!=\!-r|\!\sin{\theta}|^{2}\\
&R_{\theta\varphi\varphi}\!=\!-r^{2}\sin{\theta}\cos{\theta}
\end{eqnarray}
and
\begin{eqnarray}
&R_{t\varphi\theta}
\!=\!-r\left[\frac{q^{2}\sin{\theta}\cos{\theta}}{1-|q^{2}\sin{\theta}|^{2}}\right]\\
&R_{r\theta\theta}
\!=\!-r\left[1\!-\!\frac{\sqrt{1-q^{4}}}{1-|q^{2}\sin{\theta}|^{2}}\right]
\end{eqnarray}
where $q$ is the electric charge and where $E\!=\!m(1\!-\!q^{4})^{\frac{1}{2}}$ is the energy of the orbital. This background can be gotten for $\alpha\!=\!\alpha(\theta)$ and $\gamma\!=\!\gamma(\theta)$ with $\varepsilon\!=\!0$ as a special instance of the more general background we have built so far.

However, even more general backgrounds are possible if the constant $\varepsilon$ is allowed to be different from zero, and a general result is that it becomes possible to have a shift in energy according to $E\!=\!m(1\!-\!q^{4})^{\frac{1}{2}}\!-\!\varepsilon$ without modifying the structure of the $1S$-orbital spinor field solution.

The fact that the very same spinor may have different energy levels looks like a degeneracy. The difference with respect to the known solution is that in this modification the change is in the form of the tensorial connection.

The tensorial connection may thus encode some of the information about the energy of the spinorial field.
\section{Solutions in free case}
As we had already remarked in \cite{Fabbri:2018crr} and as we have recalled here above, some of the components of the tensorial connection may contain information about the energy of spinor fields for electronic orbitals: specifically, they give rise to a $\Delta E\!=\!-\varepsilon$ energy shift. For $\varepsilon\!>\!0$ we obtain some negative energy contribution corresponding to having an attraction in the structure of the covariant inertial force.

In the rest of the article, we are going to neglect electrodynamics; for spinors in the form of a spin-eigenstate along the third axis, rotations around the third axis have the same effect as gauge transformations: hence, in this case we can always choose to have $P_{k}\!=\!0$ identically and without spoiling the polar structure of the spinor field.

As a final restriction, we require that $\gamma$ be a constant which will be taken to be $\gamma\!=\!\pi/2$ for simplicity: then
\begin{eqnarray}
&R_{\theta tt}\!=\!R_{\varphi\theta t}\!=\!0
\end{eqnarray}
as well as
\begin{eqnarray}
&R_{\theta rr}\!=\!0
\end{eqnarray}
and
\begin{eqnarray}
&R_{r\theta\theta}\!=\!-r
\end{eqnarray}
and where the remaining components contain either the function $\alpha$ or the constant $\varepsilon$ and as such they are physical.

When they are substituted into the Dirac equation in polar form (\ref{dep1}, \ref{dep2}) we remain with
\begin{eqnarray}
&\frac{1}{r}\partial_{\theta}\alpha\!+\!\partial_{r}\beta\!=\!0\\
&\frac{2}{r}\!-\!2\varepsilon\sinh{\alpha}\!+\!\partial_{r}\ln{\phi^{2}}\!=\!0\\
&-r\partial_{r}\alpha\!-\!2\varepsilon r\cosh{\alpha}
\!+\!\partial_{\theta}\beta\!+\!2rm\cos{\beta}\!=\!0\\
&\cot{\theta}\!+\!2rm\sin{\beta}\!+\!\partial_{\theta}\ln{\phi^{2}}\!=\!0
\end{eqnarray}
which have to be worked out in order to find solutions.

One peculiar solution can be obtained with a vanishing Yvon-Takabayashi angle: for $\beta\!=\!0$ we have $\alpha\!=\!\alpha(r)$ as
\begin{eqnarray}
&\alpha'\!+\!2\varepsilon\cosh{\alpha}\!=\!2m
\end{eqnarray}
of which one solution is for
\begin{eqnarray}
&\tanh{\alpha}\!=\!-\sqrt{1\!-\!\varepsilon^{2}/m^{2}}
\end{eqnarray}
so that
\begin{eqnarray}
&\phi\!=\!Ke^{\left(-r\sqrt{m^{2}-\varepsilon^{2}}\right)}r^{-1}/\sqrt{\sin{\theta}}
\end{eqnarray}
defined for $m\!>\!\varepsilon\!>\!0$ as is quite straightforward to verify.

Therefore, the full solution is given by
\begin{eqnarray}
&s_{\theta}\!=\!r
\end{eqnarray}
and
\begin{eqnarray}
&u_{t}\!=\!m/\varepsilon\\
&u_{\varphi}\!=\!-r\sin{\theta}\sqrt{m^{2}/\varepsilon^{2}\!-\!1}
\end{eqnarray}
for
\begin{eqnarray}
&\beta\!=\!0
\end{eqnarray}
and
\begin{eqnarray}
&\phi\!=\!Ke^{\left(-r\sqrt{m^{2}-\varepsilon^{2}}\right)}r^{-1}/\sqrt{\sin{\theta}}
\end{eqnarray}
which does not have singular points, apart from the expected divergence along the polar axis, and its square is integrable, as it has exponential convergence at infinity.

In fact, its integral is given by
\begin{eqnarray}
\int\Phi dV\!=\!\frac{2\pi^{2}K^{2}}{\sqrt{m^{2}-\varepsilon^{2}}}
\end{eqnarray}
which is finite so long as condition $m\!\neq\!\varepsilon$ is respected, and we notice that also the condition $\varepsilon\!\neq\!0$ must be respected, since for $\varepsilon\!=\!0$ we would have $\phi^{2}r^{2}\sin{\theta}\!=\!K^{2}$ which is not square-integrable, as it is known since this is the solution we would have if we were in the standard situation.

Therefore, it is essential for this solution to be defined on a background with a structure richer than usual.

Notwithstanding, that the $R_{ijk}$ tensor has components less trivial than the common ones is not a stretch of the imagination because the $R_{ijk}$ tensor is naturally present for spinors since it is necessary to have tetrads to define spinor fields in the first place. To convince skeptical readers, let us convert the presented solution from the polar form back to the common language of tetrads: to do so, it is sufficient to check that for the basis of dual tetrads
\begin{eqnarray}
&\!\!\!\!e^{0}_{t}\!=\!\cosh{\alpha}\ \ \ \ e^{2}_{t}\!=\!\sinh{\alpha}\\
&\!\!\!\!e^{1}_{r}\!=\!1\\
&e^{3}_{\theta}\!=\!-r\\
&\!\!\!\!e^{0}_{\varphi}\!=\!r\sin{\theta}\sinh{\alpha}\ \ \ \ e^{2}_{\varphi}\!=\!r\sin{\theta}\cosh{\alpha}
\end{eqnarray}
and
\begin{eqnarray}
&\!\!\!\!e_{0}^{t}\!=\!\cosh{\alpha}\ \ \ \ e_{2}^{t}\!=\!-\sinh{\alpha}\\
&\!\!\!\!e_{1}^{r}\!=\!1\\
&e_{3}^{\theta}\!=\!-\frac{1}{r}\\
&\!\!\!\!e_{0}^{\varphi}\!=\!-\frac{1}{r\sin{\theta}}\sinh{\alpha}\ \ \ \ 
e_{2}^{\varphi}\!=\!\frac{1}{r\sin{\theta}}\cosh{\alpha}
\end{eqnarray}
the spin connection is given by
\begin{eqnarray}
&\Omega_{13\theta}\!=\!-1\\
&\Omega_{01\varphi}\!=\!\sin{\theta}\sinh{\alpha}\ \ \ \ \Omega_{03\varphi}\!=\!-\cos{\theta}\sinh{\alpha}\\
&\Omega_{23\varphi}\!=\!\cos{\theta}\cosh{\alpha}\ \ \ \ \Omega_{12\varphi}\!=\!\sin{\theta}\cosh{\alpha}
\end{eqnarray}
which is obviously a non-trivial spin connection although it gives rise to a Riemann curvature that is equal to zero.

For this more common way of expressing fields, an easy computation is enough to see that in chiral representation the polar spinor given by
\begin{eqnarray}
&\!\psi\!=\!\frac{K}{\sqrt{\sin{\theta}}}\frac{1}{r}e^{\left(-r\sqrt{m^{2}-\varepsilon^{2}}\right)}
e^{-i\varepsilon t}\left(\!\begin{tabular}{c}
$1$\\
$0$\\
$1$\\
$0$
\end{tabular}\!\right)
\end{eqnarray}
is an exact solution in the free case of the Dirac equation.

Consequently, the existence of some square-integrable polar solution has been established in the instance where a non-zero tensorial connection is present in a free Dirac differential field equation. The square-integrability of the polar solution is a direct consequence of the existence of the negative energy contribution, that is of the existence of the covariant attractive inertial force given as tensorial connection, for the free Dirac differential field equation.

A more general account for such non-trivial underlying background within the polar form of the Dirac differential field equation has been provided in reference \cite{Fabbri:2018crr}.
\section{Some deepening}
Having found an explicit exact solution, the following step consists in examining its main features: to begin, we recall that the non-vanishing tensor $R_{ijk}$ is given by
\begin{eqnarray}
&R_{\theta\varphi\varphi}\!=\!-r^{2}\cos{\theta}\sin{\theta}\\
&R_{r\varphi\varphi}\!=\!-r|\!\sin{\theta}|^{2}\\
&R_{r\theta\theta}\!=\!-r\\
&R_{rtt}\!=\!2\sqrt{m^{2}\!-\!\varepsilon^{2}}\\
&R_{\varphi rt}\!=\!2mr\sin{\theta}
\end{eqnarray}
giving the square invariants
\begin{eqnarray}
&\!\!\!\!R_{ac}^{\phantom{ac}c}R^{ai}_{\phantom{ai}i}
\!=\!4\left(\varepsilon^{2}\!-\!m^{2}
\!-\!\frac{2}{r}\sqrt{m^{2}\!-\!\varepsilon^{2}}
\!-\!\frac{4+|\!\cot{\theta}|^{2}}{4r^{2}}\right)\\
&\!\!R_{ijk}R^{kij}
\!=\!4\left(m^{2}\!-\!\varepsilon^{2}
\!+\!\frac{2+|\!\cot{\theta}|^{2}}{4r^{2}}\right)\\
&\!\frac{1}{2}R_{ijk}R^{ijk}
\!=\!4\left(\varepsilon^{2}
\!-\!\frac{2+|\!\cot{\theta}|^{2}}{4r^{2}}\right)
\end{eqnarray}
and derivative invariant
\begin{eqnarray}
&\nabla_{\mu}R^{\mu\nu}_{\phantom{\mu\nu}\nu}
\!=\!-4\left(\frac{1}{r}\sqrt{m^{2}\!-\!\varepsilon^{2}}\!+\!\frac{1}{4r^{2}}\right)
\end{eqnarray}
showing that the background is richer than the common situation. We also have the derivative
\begin{eqnarray}
&\nabla_{\mu}s^{\mu}\!=\!-\frac{\cot{\theta}}{r}
\end{eqnarray}
showing that the spin has a non-trivial structure too.

The energy density tensor is
\begin{eqnarray}
&T_{\nu\alpha}\!=\!\frac{1}{2}\phi^{2}R^{ij}_{\phantom{ij}\alpha}s^{k}\varepsilon_{ijk\nu}
\end{eqnarray}
which can be calculated as
\begin{eqnarray}
&T_{\varphi t}\!=\!-2\phi^{2}r\sin{\theta}\sqrt{m^{2}\!-\!\varepsilon^{2}}\\
&T_{t\varphi}\!=\!\phi^{2}\sin{\theta}\\
&T_{tt}\!=\!2\phi^{2}m
\end{eqnarray}
with an azimuthal flux of energy and density of azimuthal momentum, as well as an energy density.

Also, the energy trace would be
\begin{eqnarray}
\int TdV\!=\!m
\end{eqnarray}
if $4\pi^{4}K^{4}\!=\!(m^{2}\!-\!\varepsilon^{2})$ is chosen as obvious normalization.

Notice that the square invariants can be taken for large radial coordinate and in this case they would all be different from zero if the condition $m\!\neq\!\varepsilon\!\neq\!0$ is still respected for the constants; because $R_{ijk}$ contains all information normally contained within the connection while being a tensor we called $R_{ijk}$ tensorial connection. It is generally non-zero but it has no curvature tensor and hence it can not have any source, so that it makes sense to think at it as what encodes some sort of covariant inertial force.

We notice that in the velocity, large relevance must be attributed to the azimuthal component, measured by the function $\alpha$ both in magnitude and sign: since $\alpha$ is strictly negative, the vorticity has direction opposite to the spin.

Also notice that the constant $\varepsilon$ has a fundamental part because by lowering the mass parameter from $m$ down to an effective mass $\sqrt{m^{2}\!-\!\varepsilon^{2}}$ it essentially plays the role of a negative energy, which is what we need for localization.

In fact the constraint $\varepsilon>0$ gives an effective negative energy, as we have seen when in the previous section we have presented the example of the Coulomb potential.

The constraints given by $m\!>\!\varepsilon$ with $\alpha\!<\!0$ respectively imply that solutions behave as hyperbolic functions and with an exponential convergence to zero at infinity.

Negative energies are proper of attractive forces, which in this case are codified through the tensorial connection by the action of covariant inertial forces \cite{Fabbri:2018crr}.
\section{Conclusion}
Finding square-integrable solutions is a nice result, but it is even more impressive to think that such a result has been obtained after a relatively large number of restrictions, and we can only guess how complex other solutions may be, if we were to allow an even richer structure of the background. The restrictions we made were three: some were restrictions on the structure of the $R_{ijk}$ tensor that consisted in having some components set to zero; others were on the structure of the velocity and spin; and a final was about assuming no Yvon-Takabayashi angle at all.

Under these restrictions, we found solutions for spinors in polar form described in terms of the module having an exponential damping with radial distance: this property relied on the fact that an overall negative contribution for the energy was present. These negative energies are the consequence of a specific type of covariant attractive inertial force introduced through the tensorial connection.

From a purely mathematical perspective, the technicalities of the presentation could be summarized by stating the following: the Dirac equation necessitates of a basis of tetrad fields beside the usual spinor structure, and so we might as well take advantage of the fact that tetrads could be non-trivial to get more general spinor solutions.

From a genuinely physical point of view we might say that since it is generally impossible to vanish the tensorial connection $R_{ijk}$ in the spinorial dynamics then we might as well use them to get more general spinorial dynamics.

The special structure of the tensorial connection entails the presence of some covariantly attractive inertial force, which is a force that despite being sourceless can not be vanished, in the same way in which the spin is a covariant angular momentum, because despite having the algebraic structure of an angular momentum we can not vanish it in an arbitrary way. Both the covariant inertial force and the spin are intrinsically defined only for spinor fields.

Nevertheless, there is a difference between these two in that we have become accustomed to spin while covariant inertial forces are a recently introduced concept.

In this paper we have discussed in what way they can be used in specific cases to find well-behaved solutions of the Dirac differential field equation.


\begin{thebibliography}{40}
\bibitem{G}
M.Gasperini, \textit{Theory of Gravitational Interactions}\\ (Springer, 2017).
\bibitem{L}
P.Lounesto, \textit{Clifford Algebras and Spinors}\\ (Cambridge University Press, 2001).
\bibitem{Cavalcanti:2014wia}
R.T.Cavalcanti, ``Classification of Singular Spinor\\ Fields and Other Mass Dimension One Fermions'',\\ \textit{Int.J.Mod.Phys.D}\textbf{23}, 1444002 (2014).
\bibitem{Fabbri:2016msm}
L.Fabbri,
``A generally-relativistic gauge\\ classification of the Dirac fields'',\\ \textit{Int.J.Geom.Meth.Mod.Phys.}\textbf{13},1650078(2016).
\bibitem{h1}
D.Hestenes, ``Real Spinor Fields'',\\ \textit{J.Math.Phys.}\textbf{8}, 798 (1967).
\bibitem{Fabbri:2016laz}
L.Fabbri,
``Torsion Gravity for Dirac Fields'',\\ \textit{Int.J.Geom.Meth.Mod.Phys.}\textbf{14},1750037(2017).
\bibitem{Fabbri:2016fxt}
L.Fabbri, ``Torsion Gravity for Dirac Particles'',\\ 
\textit{Int.J.Geom.Meth.Mod.Phys.}\textbf{14}, 1750127 (2017).
\bibitem{Obukhov:2017avp}
Y.N.Obukhov, A.J.Silenko, O.V.Teryaev, ``General\\ treatment of quantum and classical spinning particles\\ in external fields'', \textit{Phys.Rev.D}\textbf{96}, 105005 (2017).
\bibitem{Fabbri:2017pwp}
L.Fabbri, ``General Dynamics of Spinors'',\\ 
\textit{Adv. Appl. Clifford Algebras}\textbf{27}, 2901 (2017).
\bibitem{Fabbri:2017xyk}
L.Fabbri, ``Spinor Fields, Singular Structures, Charge\\ 
Conjugation, ELKO and  Neutrino Masses'',\\
\textit{Adv.Appl.Clifford Algebras}\textbf{28}, 7 (2018).
\bibitem{Fabbri:2018crr}
L.Fabbri, ``Covariant inertial forces for spinors'',\\ 
\textit{Eur.J.Phys.C}\textbf{78}, 783 (2018).
\end{thebibliography}
\end{document}